\documentclass[12pt]{article}
\usepackage{amssymb}
\usepackage{epsf}

\begin{document}
\begin{center}
{\bf \Large From Time-symmetric Microscopic Dynamics to
 Time-asymmetric Macroscopic Behavior: An Overview}

\vskip15pt

Joel L. Lebowitz

Departments of Mathematics and Physics

Rutgers, The State University

Piscataway, New Jersy

lebowitz@math.rutgers.edu

\end{center}
\vskip20pt

\noindent {\bf Abstract} Time-asymmetric behavior as embodied in the second
law of thermodynamics is observed in {\it individual macroscopic} systems.
It can be understood as arising naturally from time-symmetric microscopic
laws when account is taken of a) the great disparity between microscopic
and macroscopic scales, b) a low entropy state of the early universe, and
c) the fact that what we observe is the behavior of systems coming from
such an initial state---not all possible systems.  The explanation of the
origin of the second law based on these ingredients goes back to Maxwell,
Thomson and particularly Boltzmann.  Common alternate explanations, such as
those based on the ergodic or mixing properties of probability
distributions (ensembles) already present for chaotic dynamical systems
having only a few degrees of freedom or on the impossibility of having a
truly isolated system, are either unnecessary, misguided or misleading.
Specific features of macroscopic evolution, such as the diffusion equation,
do however depend on the dynamical instability (deterministic chaos) of
trajectories of isolated macroscopic systems.

The extensions of these classical notions to the quantum world is in many
ways fairly direct.  It does however also bring in some new problems.
These will be discussed but not resolved.

\section{ Introduction}

Let me start by stating clearly that I am not going to discuss here---much
less claim to resolve---the many complex issues, philosophical and
physical, concerning the nature of time, from the way we perceive it to the
way it enters into the space-time structure in relativistic theories.  I
will also not try to philosophize about the ``true'' nature of probability.
My goal here, as in my previous articles \cite{1, 2} on this subject, is much
more modest.\footnote{The interested reader may wish to look
at the three book reviews of which are contained in [1e], [1f].  These
books attempt to deal with some fundamental questions about time.  As
for the primitive notion of probability I have in mind something
like this: the probability that when you next check your mail box you will
find a package with a million dollars in it is very small,
c.f. section 3.} I will take (our
everyday notions of) space, time and probability as primitive undefined
concepts and try to clarify the many conceptual and mathematical problems
encountered in going from a time symmetric Hamiltonian microscopic dynamics
to a time asymmetric macroscopic one, as given for example by the diffusion
equation.  I will also take it for granted that every bit of macroscopic
matter is composed of an enormous number of quasi-autonomous units, called
atoms (or molecules).

The atoms, taken to be the basic entities making up these macroscopic
objects, will be simplified to the point of caricature: they will be
treated, to quote Feynman \cite{3}, as ``little particles that move around in
perpetual motion, attracting each other when they are a little distance
apart, but repelling upon being squeezed into one another.''  This
crude picture of atoms (a refined version of that held by some ancient
Greek philosophers) moving according to non-relativistic classical
Hamiltonian equations contains the essential qualitative and even
quantitative ingredients of macroscopic irreversibility.  To accord
with our understanding of microscopic reality it must, of course, be
modified to take account of quantum mechanics.  This raises further
issues for the question of irreversibility which will be discussed in
section 9.

Much of what I have to say is a summary and elaboration of the work
done over a century ago, when the problem of reconciling time asymmetric
macroscopic behavior with the time symmetric microscopic dynamics became a
central issue in physics.  To quote from Thomson's (later Lord Kelvin)
beautiful and highly recommended 1874 article \cite{4}, \cite{5} ``The essence
of Joule's discovery is the subjection of physical [read thermal] phenomena
to [microscopic] dynamical law.  If, then, the motion of very particle of
matter in the universe were precisely reversed at any instant, the course
of nature would be simply reversed for ever after.  The bursting bubble of
{}foam at the foot of a waterfall would reunite and descend into the water
\dots .  Physical processes, on the other hand, are irreversible: for
example, the friction of solids, conduction of heat, and diffusion.
Nevertheless, the principle of dissipation of energy [irreversible
behavior]  is compatible with a
molecular theory in which each particle is subject to the laws of abstract
dynamics.''

\subsection{Formulation of Problem}

{}Formally the problem considered by Thomson in the context of Newtonian
theory, the ``theory of everything'' at that time, is as follows: The complete
microscopic (or micro) state of a classical  system of $N$
particles is represented by a point $X$  in its phase  space $\Gamma$, $X =
({\bf r}_1, {\bf p}_1, {\bf r}_2,  {\bf p}_2, ...,  {\bf r}_N, {\bf p}_N)$,
${\bf r}_i$ and  ${\bf p}_i$ being the position and momentum (or
velocity) of the $i$th particle.  When the system is isolated its evolution is governed  by
Hamiltonian dynamics with some
specified Hamiltonian $H(X)$ which we will assume for simplicity to be
an even function of the momenta.  Given $H(X)$, the microstate $X(t_0)$, at
time $t_0$, determines the  microstate  $X(t)$  at  all future and
past times $t$ during which the system  will be or was isolated: $X(t) =T_{t-t_0}X(t_0)$.    Let  $X(t_0)$ and  $X(t_0+\tau)$, with $\tau$
positive,   be  two  such  microstates.    Reversing   (physically  or
mathematically)   all  velocities at  time   $t_0+\tau$,  we obtain   a new
microstate.  If we now follow the  evolution for another interval $\tau$ we
{}find that the new microstate  at time $t_0 +  2\tau$ is just $RX(t_0)$, the
microstate $X(t_0)$ with all   velocities reversed:  $RX = ({\bf  r}_1,-{\bf
p}_1,  {\bf  r}_2,-{\bf  p}_2, ...,{\bf r}_N,-{\bf p}_N)$.  Hence if there  is an evolution, i.e.\ a
trajectory $X(t)$, in which some property of the system, specified by a
{}function $f(X(t))$, behaves in a certain way as $t$  increases, then if $f(X) =f(RX)$ there  is  also a  trajectory in  which the property  evolves in the
time reversed direction.  Thus, for example, if  particle densities get more
uniform as time increases, in a way described by the diffusion equation, then since the density
profile is the same for $X$ and $RX$ there is also an evolution in which the
density gets more nonuniform.  So why is one type of evolution,
the one consistent with an entropy increase in accord with the
``second law'', common and the other never seen?  The difficulty is illustrated
by the impossibility of time
ordering of the snapshots in {Fig.\ 1} using {\it solely} the microscopic
dynamical laws: the above time symmetry implies that if (a, b, c, d)
is a possible ordering so is (d, c, b, a).

\epsfysize=2in
\begin{figure} [t]
\centerline{\epsffile{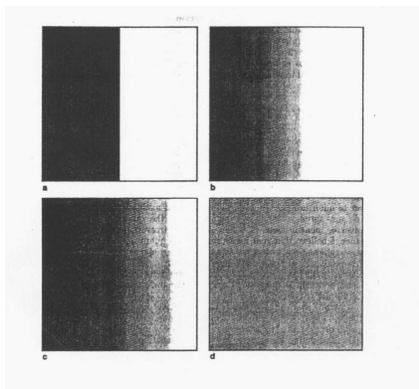}}
\caption{\it A sequence of ``snapshots'', a, b, c, d taken at times
$t_a, t_b, t_c, t_d$,
each representing a macroscopic state of a system, say a fluid with two
``differently colored'' atoms or a solid in which the shading indicates
the local temperature.  How would you order this sequence in time?}
\end{figure}

\subsection{Resolution of Problem}

The explanation of this apparent paradox, due to Thomson, Maxwell and
Boltzmann, as described in references \cite{1}--\cite{17}, which I will
summarize in this article, shows that {\it not only is
there no conflict} between reversible microscopic laws and irreversible
macroscopic behavior, but, as clearly pointed out by Boltzmann in his later
writings\footnote{Boltzmann's early writings on the subject are
sometimes unclear, wrong, and even contradictory.  His later writings,
however, are generally very clear and right on the money (even if a bit
verbose for Maxwell's taste, c.f.\ \cite{8}.)   The presentation here is not
intended to be historical.}, {\it there are extremely strong reasons
to expect the latter from the former}.  These reasons involve
several interrelated ingredients which together provide the required
distinction between microscopic and macroscopic variables and explain the
emergence of definite time asymmetric behavior in the evolution of the
latter despite the total absence of such asymmetry in the dynamics of the
former.  They are: a) the great disparity between microscopic and
macroscopic scales, b) the fact that the events we observe in our
world are determined not only by the microscopic dynamics, but also by
the initial conditions of our system, which, as we shall see later, in
section 6, are very much related to the initial conditions of our
universe, and c) the fact that it is not
every microscopic state of a macroscopic system that will evolve in
accordance with the entropy increase predicted by the second law, but
only the ``majority'' of such states---a majority which however becomes so
overwhelming when the number of atoms in the system becomes very large that
irreversible behavior becomes effectively a certainty.  
To make the last statement complete we shall have to specify the
assignment of weights, or probabilities, to
different microstates consistent with a given macrostate.  Note, however,
that since we are concerned with events which have overwhelming
probability, many different assignments are equivalent and there is no
need to worry about them unduly.  
There is however, as we
shall see later, a ``natural'' choice based on phase space volume (or
dimension of Hilbert space in quantum mechanics).  These considerations enabled Boltzmann
to define the entropy of a macroscopic system in terms of its
microstate and to relate its
change, as expressed by the second law, to the evolution of the system's
microstate.  We detail below how the above explanation works by
describing first how to specify the macrostates of a macroscopic
system.  It is in the time evolution of these macrostates that we
observe irreversible behavior \cite{1}--\cite{17}.

\subsection{Macrostates}

To describe the macroscopic state of a system of $N$ atoms in a box $V$,
say $N \gtrsim  10^{20}$, with the volume of $V$, denoted by $|V|$, satisfying
$|V|\gtrsim N l^3$, where $l$ is a typical atomic length
scale, we make use of  a much cruder description than that provided
by the microstate $X$, a point in the $6N$ dimensional phase space $\Gamma
= V^N
\otimes \mathbb{R}^{3N}$.  We shall denote by $M$ such a macroscopic
description or macrostate. As an example we may take $M$ to 
consist of the specification, to within a given accuracy, of the energy
and number of particles in each half of the box $V$.  A more refined macroscopic description would divide $V$ into $K$ cells, where
$K$ is large but still $K << N$, and specify the number of particles, the
momentum, and the amount
of energy in each cell, again with some
tolerance.   For many purposes it is convenient to consider cells
which are small on the macroscopic scale yet contain many atoms.  
This leads to a description of the macrostate in terms of smooth
particle, momentum and energy densities, such as those used in the
Navier-Stokes equations  \cite{18}, \cite{19}.  An even
more refined description is obtained by considering a smoothed 
out density $f({\bf r}, {\bf p})$ in the six-dimensional position and
momentum space such as enters the Boltzmann equation for dilute
gases \cite{17}.  (For dense systems this needs to be
supplemented by the positional
potential energy density; see footnote {$d$} and reference \cite{2}
{}for details.)

Clearly $M$ is determined by $X$ (we will thus write $M(X)$) but there are many $X$'s (in fact a
continuum) which 
correspond to the same
$M$.  Let $\Gamma_M$ be the region in $\Gamma$ consisting of all
microstates $X$ corresponding to a given macrostate $M$ and denote by
$|\Gamma_M| = (N! h^{3N})^{-1} 
\int_{\Gamma_M} \Pi^N_{i=1} d{\bf r}_i d{\bf p}_i$, its symmetrized
$6N$ dimensional Liouville volume (in units of $h^{3N}$).    

\subsection{Time Evolution of Macrostates: An Example}

Consider a situation in which a gas of $N$ atoms with energy $E$ 
(with some tolerance) is
initially confined by a partition to the left half of of the box $V$,
and suppose that this constraint is
removed at time $t_a$, see Fig. 1.  
The phase space volume available to the system for times 
$t>t_a$ is then
{}fantastically enlarged\footnote{If the system
contains 1 mole of gas then the volume ratio of the unconstrained phase
space region to the constrained one is
{}far larger than the ratio of the volume of the known universe to the volume
of one proton.} compared to what it was initially, roughly by a factor
of $2^N$.  

Let us now consider the macrostate of this gas as given by
$M=\left({N_L \over N} , {E_L \over E}\right)$, the fraction of
particles and energy in the left half of $V$ (within some small
tolerance).  The macrostate at time $t_a, M=(1, 1)$,
will be denoted by $M_a$.  The phase-space region $|\Gamma| =
\Sigma_E$, 
available to the system for $t> t_a$, i.e., the
region in which  $H(X) \in (E, E + \delta E),  \delta E << E$,
will contain new macrostates,
corresponding to various fractions of particles and energy in the left half of the
box, with phase space volumes very large compared to the initial phase
space volume available to the system.  We can then expect (in the absence
of any obstruction, such as a hidden conservation law) that as the phase
point $X$ evolves under the unconstrained dynamics and explores the newly
available regions of phase space, it will with very high probability enter
a succession of new macrostates $M$ for which $|\Gamma_{M}|$ is
increasing.  The set of all the phase points $X_t$, which at time $t_a$ were in
$\Gamma_{M_a}$, forms a region $T_t\Gamma_{M_a}$ whose volume
is, by Liouville's Theorem, equal to $|\Gamma_{M_a}|$.
The shape of $T_t\Gamma_{M_a}$ will however change with $t$ and as
$t$ increases $T_t\Gamma_{M_a}$ will increasingly be contained in 
regions $\Gamma_M$ corresponding to macrostates with larger and 
larger phase space volumes $|\Gamma_M|$. 
This will continue until almost all the phase points initially in 
$\Gamma_{M_a}$ are contained in $\Gamma_{M_{eq}}$, with $M_{eq}$
the system's unconstrained macroscopic
equilibrium state.    
This is the state in which approximately half the particles and half
the energy will be
located in the left half of the box, $M_{eq} = 
({1\over 2}, {1 \over 2})$ i.e. 
$N_L /N$ and $E_L/ E$ will each
be in an interval $\left({1 \over 2} -
\epsilon, {1 \over 2} + \epsilon\right)$, $N^{-1/2} << \epsilon << 1$.

$M_{eq}$ is characterized, in fact defined, by the fact that 
it is the unique macrostate, among all the $M_\alpha$, for which
$|\Gamma_{M_{eq}}| / |\Sigma_E| \simeq 1$, where $|\Sigma_E|$ is the
total phase space
volume available under the energy constraint $H(X) \in (E, E + \delta
E)$.  (Here the symbol
$\simeq$ means equality when $N \to \infty$.)   That there exists a
macrostate containing almost all of the microstates in $\Sigma_E$ is a
consequence of the law of large numbers \cite{20}, \cite{18}.  The fact that 
$N$ is enormously large for macroscope systems
is absolutely critical for the existence of thermodynamic equilibrium
states for any reasonable definition of macrostates, e.g. for any
$\epsilon$, in the above example
such that $N^{-1/2} << \epsilon << 1$.  Indeed thermodynamics does 
not apply (is even meaningless) for isolated systems containing 
just a few particles, c.f. Onsager
\cite{21} and Maxwell quote in the next section \cite{22}.  Nanosystems are interesting and important intermediate cases
which I shall however not discuss here; see related discussion about
computer simulations in footnote {$e$}.  

After reaching $M_{eq}$ we will (mostly) see only small fluctuations in
$N_L(t) / N$ and $E_L(t) / E$,  about
the value ${1 \over 2}$: typical
{}fluctuations in $N_L$ and $E_L$
being of the order of the square root of the number of
particles involved \cite{18}.  
(Of course if the system remains isolated long enough
we will occasionally also see a return to the initial macrostate---the expected
time for such a Poincar\'e recurrence is however much longer than the 
age of the universe and so is of
no practical relevance when discussing the approach to equilibrium of
a macroscopic system \cite{6}, \cite{8}.)  

As already noted earlier the scenario in which $|\Gamma_{M(X(t))}|$
increase with time for the $M_a$ shown in Fig.1
cannot be true for all
microstates $X\subset \Gamma_{M_a}$.  There will of necessity be $X$'s
in $\Gamma_{M_a}$ which will evolve for a certain amount of time
into microstates $X(t)\equiv X_t$ such that
$|\Gamma_{M(X_t)}|<|\Gamma_{M_a}|$, e.g. microstates $X\in
\Gamma_{M_a}$ which have all velocities directed away from the
barrier which was lifted at $t_a$.  What is true however is that
the subset $B$ of such ``bad'' initial states has a phase space volume
which is very very small compared to that of $\Gamma_{M_a}$.  This is
what I mean when I say that entropy increasing behavior is {\it
typical}; a more extensive discussion of typicality is given later.

\section{Boltzmann's Entropy}

The end result of the time evolution in
the above example, that of the fraction of particles and energy
becoming and
remaining essentially equal in the
two halves of the container when $N$ is large enough 
(and `exactly equal' when $N \to\infty$), is of course what is predicted by the
second law of thermodynamics.  According to this law the  final state of an
isolated system with specified constraints on the energy, volume, 
and mole number is one in which the entropy, a measurable macroscopic
quantity of equilibrium systems, defined on a purely operational level by
Claussius, has its maximum.  (In practice one also fixes additional
constraints, e.g. the chemical combination of
nitrogen and oxygen to form complex molecules is ruled out when
considering, for example, the dew point of air in the `equilibrium'
state of air at normal temperature and pressure, c.f.\ \cite{21}.
There are, of course, also very long lived metastable states,
e.g. glasses, which one can, for many purposes, treat as equilibrium states
even though their entropy is not maximal.  I will ignore these
complications here.)
 In our example this thermodynamic entropy would be given by 
$S= V_Ls\left({N_L\over V_L}, {E_L\over V_L}\right) + V_R s\left(
{N_R\over V_R}, {E_R\over V_R}\right)$ defined  for all
equilibrium states in separate boxes $V_L$ and $V_R$ with given values of
$N_L, N_R, E_L, E_R$.  When $V_L$ and $V_R$ are united to form $V, S$ is 
maximized  subject to the constraint of $E_L + E_R = E$ and of  $N_L +
N_R = N$.  

It was Boltzmann's great insight to  connect the second law with
the above phase space volume considerations by making the observation that for a
dilute gas $\log |\Gamma_{M_{eq}}|$ is proportional, up to terms negligible
in the  size of   the system,  to  the  thermodynamic entropy of  Clausius.
Boltzmann then extended his insight about the  relation between thermodynamic entropy  and
$\log |\Gamma_{M_{eq}}|$ to all macroscopic systems; be they gas,
liquid or  solid.   This provided for  the first  time a microscopic
definition of the operationally  measurable
entropy of macroscopic systems in {\it equilibrium}.

Having made this connection Boltzmann then generalized it to define an
entropy also for macroscopic systems not in equilibrium.  That is, he
associated with each microscopic state $X$ of a macroscopic system a
number $S_B$ which depends only on $M(X)$ given, up to multiplicative and additive constants (which can
depend on $N$), by
$$
S_B(X) = S_B (M(X))    \eqno(1a)
$$
with 
$$
S_B(M) = k \log|\Gamma_{M}|,    \eqno(1b)
$$
which, following O. Penrose \cite{13}, I shall call the Boltzmann entropy
of a classical system: $|\Gamma_M|$ is defined in section (1.3).
 N. B.  I have deliberately written (1) as two equations to
emphasize their logical independence which will be useful for the
discussion of quantum systems in section 9.

 Boltzmann then used phase space arguments, 
like those given above, to explain
(in agreement with the ideas of Maxwell and Thomson) the observation,
embodied in the second law of thermodynamics, that when a constraint is
lifted, an isolated macroscopic system will evolve toward a state with
greater entropy.\footnote{When $M$ specifies a state of
local equilibrium, $S_B(X)$ agrees up to negligible terms, with the
``hydrodynamic entropy''.  For systems far from equilibrium the appropriate
definition of $M$ and thus of $S_B$ can be more problematical.  For a dilute
gas (with specified kinetic energy and negligible potential energy) in
which $M$ is specified by the smoothed empirical density $f({\bf r}, {\bf v})$ 
of atoms in the
six dimensional position and velocity space, $S_B(X) = - k\int f({\bf r},
{\bf v}) \log f({\bf r}, {\bf v}) d{\bf r} d{\bf v}$ (see end of
Section 4).  This identification
is, however, invalid when the potential energy is not negligible and
one has to add to $f({\bf r}, {\bf v})$ also information about the energy
density.  This is discussed in detail in \cite{2}.   Boltzmann's
{}famous $H$ theorem derived from his eponymous equation for dilute
gases is thus an expression of the second law applied to the
macrostate specified by $f$.  It was also argued in \cite{2} that such an 
$H$ theorem must hold whenever there is a 
deterministic equation for the macrovariables of an isolated system.}
In effect Boltzmann argued that due to the large differences in the
sizes of $\Gamma_M$, $S_B(X_t) = k \log |\Gamma_{M(X_t)}|$ will {\it typically}
increase in a way which {\it explains} and describes
qualitatively the evolution towards equilibrium of macroscopic
systems. 

These very large differences in the values of $|\Gamma_M|$ for different $M$
come from the very large number of particles (or degrees of freedom)
which contribute, in an (approximately) additive way, to the
specification of macrostates.  This is also what gives rise to typical or
almost sure behavior. Typical, as used here, means that the set of microstates corresponding to a
given macrostate $M$ for which the evolution leads to a macroscopic
increase (or non-decrease) in the Boltzmann entropy during some fixed
macroscopic time period $\tau$
occupies a subset of $\Gamma_M$ whose Liouville volume is a fraction of
$|\Gamma_M|$ which goes very rapidly (exponentially) to one as the number
of atoms in the system increases.  The fraction of ``bad''
microstates, which lead to an entropy decrease, thus goes to zero as
$N\to \infty$.

Typicality is what distinguishes macroscopic irreversibility 
{}from the weak approach to
equilibrium of probability distributions (ensembles) of systems
with good ergodic properties having only a few degrees of freedom,
e.g. two hard spheres in a cubical box.  While the former is
manifested in a
typical evolution of a single macroscopic system the latter does not
correspond to any appearance of time asymmetry in the evolution of an
individual system.  Maxwell makes clear the importance of the separation
between microscopic and macroscopic scales when he writes \cite{22}: 
``the second law is drawn from our experience of bodies 
consisting of an immense number
of molecules.  ... it is continually being violated, ..., in any
sufficiently small group of molecules ... .  As the number ... is increased
... the probability of a measurable variation ... may be regarded as
practically an impossibility.''  This is also made very clear by
Onsager in \cite{21} 
and should be contrasted with the confusing statements found in many 
books that thermodynamics can be applied to a single isolated particle 
in a box, c.f. footnote $i$.

On the other hand, because of the exponential increase of the phase space
volume with particle number, even a system with only a few hundred
particles, such as is commonly used in 
molecular dynamics computer simulations, will, when
started in a nonequilibrium `macrostate' $M$, with `random' $X \in
\Gamma_M$, appear to behave like a macroscopic
system.\footnote{After all, the likelihood of hitting, in
the course of say one thousand tries, something which has probability of
order $2^{-N}$ is, for all practical purposes, the same, whether $N$ is a
hundred or $10^{23}$.  Of course the fluctuation in $S_B$ both along
the path towards equilibrium and in equilibrium will be larger when
$N$ is small, c.f. [2b].}  This will be so even when 
integer arithmetic is
used in the simulations so that the system behaves as a truly isolated one;
when its velocities are reversed the system retraces its steps until it
comes back to the initial state (with reversed velocities), after which it
again proceeds (up to very long Poincare recurrence times) in the typical
way, see section 5 and Figs.\ 2 and 3 taken from \cite{23} and \cite{24}.

\epsfysize=2in
\begin{figure}[t]
\centerline{\epsffile{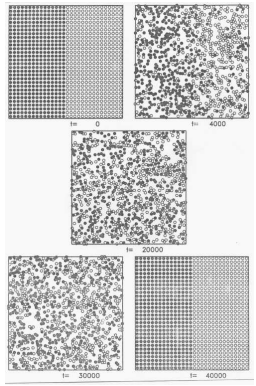}}
\caption{\it Time evolution of a system of 900 particles all  
interacting
via the same cutoff Lennard-Jones pair potential using integer
arithmetic.  Half of the particles are colored white, the other half black.
All velocities are reversed at $t=20,000$.  The system then retraces its
path and the initial state is fully recovered.  From Levesque and Verlet,
Ref.\ \cite{23}.} 
\end{figure}

\epsfysize=2in
\begin{figure}[t]
\centerline{\epsffile{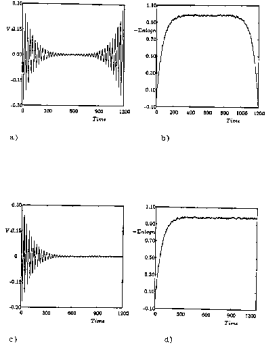}}

\caption{\it Time evolution of a reversible cellular automaton lattice gas using
integer arithmetic.  Figures a) and c) show the mean velocity, figures b)
and d) the entropy.  The mean velocity decays with time and the entropy
increases up to $t=600$ when there is a reversal of all velocities.  The
system then retraces its path and the initial state is fully recovered in
{}figures a) and b).  In the bottom figures there is a small error in the
reversal at $t=600$.  While such an error has no appreciable effect on the
initial evolution it effectively prevents any recovery of the initial
state.  The entropy, on the scale of the figure, just remains at its
maximum value.  This shows the instability of the reversed path.  From
Nadiga et al. Ref.\ \cite{24}.}
\end{figure}

We might take as a summary of such insights in the late part of the nineteenth
century the statement by Gibbs \cite{25} quoted by Boltzmann (in a German
translation) on the cover of his book {\it Lectures on Gas Theory II}:
\cite{7}, ``In
other words, the impossibility of an uncompensated decrease of entropy
seems to be reduced to an improbability.''

\section{The Use of Probabilities}

As already noted, typical here means overwhelmingly probable with 
respect to a measure which assigns (at least
approximately) equal weights to regions of equal phase space volume within
$\Gamma_M$ or, loosely speaking, to different microstates consistent with the
``initial'' macrostate $M$.  (This is also what was meant earlier by the
`random' choice of an initial $X \in \Gamma_M$ in the computer
simulations.)  In fact, any mathematical statement about probable or
improbable behavior of a physical system has to refer to some agreed upon
measure (probability distribution).  It is, however, very hard (perhaps
impossible) to formulate precisely what one means, as a statement about the
real world, by an assignment of exact numerical values of probabilities (let alone rigorously justify
any particular one) in our context.  It is therefore not so surprising that
this use of probabilities, and particularly the use of typicality for
explaining the origin of the apparently deterministic second law, was very
difficult for many of Boltzmann's contemporaries, and even for some people
today, to accept. (Many text books on statistical mechanics are
unfortunately either silent or confusing on this very important point.) This was clearly very frustrating to Boltzmann as it is also
to me, see [1b, 1c].  I have not found any better way of expressing this frustration than
Boltzmann did when he wrote, in his second reply to Zermelo in 1897 \cite{6} ``The
applicability of probability theory to a particular case cannot of course
be proved rigorously. ...  Despite this, every insurance company relies on
probability theory. ... It is even more valid [here], on account of the
huge number of molecules in a cubic millimetre...  The assumption that
these rare cases are not observed in nature is not strictly provable (nor
is the entire mechanical picture itself) but in view of what has been said
it is so natural and obvious, and so much in agreement with all experience
with probabilities ... [that] ... {\it It is completely incomprehensible to
me} [my italics] how anyone can see a refutation of the applicability of
probability theory in the fact that some other argument shows that
exceptions must occur now and then over a period of eons of time; for
probability theory itself teaches just the same thing.''

The use of probabilities in the Maxwell-Thomson-Boltzmann explanation of
irreversible macroscopic behavior is as Ruelle notes ``simple but
subtle'' \cite{14}.  They introduce into the laws of nature notions of
probability, which, certainly at that time, were quite alien to the
scientific outlook.  Physical laws were supposed to hold without any
exceptions, not just almost always and indeed no exceptions were (or are)
known to the second law as a statement about the actual behavior of
isolated macroscopic systems;
nor would we expect any, as Richard Feynman \cite{15} 
rather conservatively says, ``in a million years''.  The reason for this,
as already noted, is that for a macroscopic 
system the fraction (in terms of the Liouville volume) of the
microstates in a macrostate $M$
{}for which the evolution leads to
macrostates $M'$
with $S_B(M')\geq  S_B(M)$ is so close to one (in terms of their
Liouville volume) that such behavior is exactly what should be seen to
``always'' happen.  
Thus in Fig.\ 1 the sequence going from left to
right is typical for a phase point in $\Gamma_{M_a}$ while the one going
{}from right to left has probability approaching zero with respect to a
uniform distribution in $\Gamma_{M_d}$, when $N$, the number of
particles (or degrees of freedom) in the system, is sufficiently large.
The situation can be quite different when $N$ is small as noted in the
last section: see Maxwell quote there.

Note that Boltzmann's explanation of why $S_B(M_t)$ is never seen to
 decrease with $t$ does not really require the assumption that
 over very long periods of time a macroscopic system should be found in
 every region $\Gamma_M$, i.e.\ in every macroscopic states $M$,
{}for a fraction of time {\it exactly} equal to the ratio of $|\Gamma_M|$ to
 the total phase space volume specified by its energy.  This latter behavior,
 embodied for example in Einstein's formula 
$$
Prob  \{M\} \sim \exp  [S_B (M) - S_{eq}]
\eqno(2)
$$
{}for fluctuation in equilibrium sytems, with probability there
interpreted as the fraction of time which such a system will spend in
$\Gamma_M$, can be considered as a mild form of the ergodic hypothesis, mild
because it is only applied to those regions of the phase space
representing macrostates $\Gamma_M$.  This seems very plausible in the absence
of constants of the motion which decompose the energy surface into regions
with different macroscopic states.  It appears even more reasonable when
we take into account the lack of perfect isolation in practice which will
be discussed later.  Its implication for small fluctuations from
equilibrium is certainly consistent with observations.  In particular when
the exponent in (2) is expanded in a Taylor series and only quadratic
terms are kept, we obtain a Gaussian distribution for normal (small)
{}fluctuations from equilibrium.  Eq.(2) is in fact one of the main 
ingredients of Onsager's reciprocity relations for transport processes
in systems close to equilibrium \cite{26}. 

The usual ergodic hypothesis, i.e. that the fraction of time spent by a
trajectority $X_t$ in any region $A$ on the energy surface $H(X) = E$
is equal to the fraction of the volume occupied by $A$, also seems like
a natural assumption for macroscopic systems.  It is however not necessary for identifying
equilibrium properties of macroscopic systems with those obtained
{}from the microcanonical ensemble;  see Section 7.  Neither is it in any
way sufficient for explaining the approach to equilibrium observed in
real systems: the time scales are entirely different.

It should perhaps be emphasized again here that an important
ingredient in the whole picture of the time evolution of macrostates
described above is the constancy in time of the Liouville volume of 
sets in the phase space $\Gamma$ as they evolve under the Hamiltonian 
dynamics (Liouville's Theorem).  Without this invariance the
connection between phase space
volume and probability would be impossible or at least very problematic.

{}For a somewhat different viewpoint on the issues discussed in this
section the reader is referred to Chapter IV in [13].

\section{Initial Conditions}

Once we accept the statistical explanation of why macroscopic systems
evolve in a manner that makes $S_B$ increase with time, there remains the
nagging problem (of which Boltzmann was well aware) of what we mean by
``with time'': since the microscopic dynamical laws are symmetric, the two
directions of the time variable are {\it a priori} equivalent and thus must
remain so {\it a posteriori}.  This was well expressed by Schr{\"o}dinger
\cite{27}.  ``First, my good friend, you state that the two directions of your
time variable, from $-t$ to $+t$ and from $+t$ to $-t$ are a priori
equivalent.  Then by fine arguments appealing to common sense you show
that disorder (or `entropy') must with overwhelming probability
increase with time.  Now, if you please, what do you mean by
`with time'?  Do you mean in the direction $-t$ to $+t$?  But if your
interferences are sound, they are equally valid for the direction $+t$
to $-t$.  If these two directions are equivalent a priori, then they
remain so a posteriori.  The conclusions can never invalidate the
premise.  Then your inference is valid for both directions of time,
and that is a contradiction.''

In terms of our Fig.\ 1 this question may be put as 
{}follows:\footnote{The reader should think of 
{}Fig.\ 1 as representing energy density in a solid: the darker the 
hotter.  The time evolution 
of the macrostate will then be given by the heat (diffusion) equation.} 
why can we use phase space arguments (or time asymmetric diffusion type
equations) to predict the macrostate at time $t$ of an {\it isolated} 
system whose macrostate at time $t_b$ is $M_b$, in the 
{}future, i.e.\ for $t > t_b$, but not in the past, i.e.\ for 
$t < t_b$?  After all, if the
macrostate $M$ is invariant under velocity reversal of all the atoms, then
the same prediction should apply equally to $t_b + \tau$ and $t_b -\tau$.
A plausible answer to this question is to assume that the nonequilibrium
macrostate $M_b$ had its origin in an even more nonuniform
macrostate $M_a$, prepared by some experimentalist at some earlier time
$t_a < t_b$ (as is indeed the case in Figure 1) and that for states thus prepared we can apply our
(approximately) equal a priori probability of microstates argument, i.e.\
we can assume its validity at time $t_a$.  But what about events on the sun
or in a supernova explosion where there are no experimentalists?  And what,
{}for that matter, is so special about the status of the experimentalist?
Isn't he or she part of the physical universe?

Before trying to answer these ``big'' questions let us consider
whether the assignment of equal probabilities for $X \in \Gamma_{M_a}$ at
$t_a$ permits the use of an equal probability distribution of $X \in
\Gamma_{M_b}$ at time $t_b$ for predicting macrostates at times $t >
t_b > t_a$
when the system is isolated for $t > t_a$.  Note that those microstates in
$\Gamma_{M_b}$ which have come from $\Gamma_{M_a}$ through the time
evolution during the time interval from $t_a$ to $t_b$ make up a set
$\Gamma_{ab}$ whose volume $|\Gamma_{ab} |$ is by Liouville's theorem
at most equal
\footnote{$|\Gamma_{ab}|$ may be strictly
less than $|\Gamma_{M_a}|$ because some of the phase points in
$\Gamma_{M_a}$ may not go into $\Gamma_{M_b}$.  There will be
approximate equality when $M_a$ at time $t_a$, determines $M_b$ at
time $t_b$: say via the diffusion equation for the energy density.  This corresponds to the ``Markov case'' discussed in
\cite{13}.  There are of course situations where the macrostate at
time $t$,
depends also (weakly or even strongly) on the whole history of $M$ in some
time interval prior to $t$, e.g.\ in materials with memory.
The second law certainly holds also for these cases - with the
appropriate definition of $S_B$, obtained in many case by just
refining the description so that the new macro variables
{}follow autonomous laws \cite{13}.}  to $|\Gamma_{M_a}|$; which, as already
discussed before, is only a very
small fraction of the volume of $\Gamma_{M_b}$.
Thus we have to show that the overwhelming majority of phase points in
$\Gamma_{ab}$ (with respect to Liouville measure on $\Gamma_{ab}$), have {\it future}
macrostates like those typical of $\Gamma_b$---while still being very
special and unrepresentative of $\Gamma_{M_b}$ as far as their {\it past}
macrostates are concerned.  This
property is explicitly proven by Lanford in his derivation of the Boltzmann
equation (for short times) \cite{17}, and is part of the derivation of
hydrodynamic equations \cite{18}, \cite{19}; see also \cite{28}.

To see intuitively the origin of this property we note that for systems
with realistic interactions the phase space region $\Gamma_{ab}\subset
\Gamma_{M_b}$ will be so convoluted
as to {\it appear} uniformly smeared out in $\Gamma_{M_b}$. It is therefore
reasonable that the future behavior of the system, as far as macrostates
go, will be unaffected by their past history.  It would of course be nice
to prove this in all cases, ``thus justifying'' (for practical purposes) the
{}factorization or ``Stosszahlansatz'' assumed by Boltzmann in deriving his
dilute gas kinetic equation for all times $t > t_a$, not only for the short
times proven by Lanford \cite{17}.   However, our mathematical abilities are equal to
this task only in very simple models such as the Lorentz gas in a
Sinai billiard.  This model describes the evolution of a macroscopic
system of independent particles moving according to Newtonian dynamics
in a periodic array of scatterers.  For this system one can actually
derive a diffusion equation for the macroscopic density profile
$n({\bf r}, t)$ from the Hamiltonian dynamics \cite{18};  see Section 8.

This behavior can also be seen explicitely in a many particle system,
each of which evolves independently  according to the reversible and area
preserving baker's transformation (which can be thought of as a toy
version of the above case)  see \cite{29}.  Here the phase space $\Gamma$
{}for $N$ particles is the $2N$ dimensional unit hypercube,
i.e. $X$ corresponds to specifying $N$-points $(x_1, y_1, \ldots,
x_N, y_N)$ in the unit square.  The discrete time evolution is given
by
$$
(x_i, y_i) \rightarrow \left\{ \begin{array}{c}
(2x_i, {1\over 2} y_i), 0 \leq x_i < {1\over
2}\\
  (2x_i -1, {1\over 2} y_i + {1\over 2}), x_i \leq {1\over 2} <
1\end{array}\right. .
$$

Dividing the unit square into
$4^k$ little squares $\delta_\alpha, \alpha = 1, \ldots, K, K=4^k$,
of side lengths $2^{-k}$, we define the macrostate $M$ by giving the
{}fraction of particles $p_\alpha = (N_\alpha/N)$
in each $\delta_\alpha$ within some tolerence.  The Boltzmann entropy
is then given, using (1) and setting $k=1$, by
$$
S_B=\sum_\alpha  \log \left[{\delta^{N_\alpha}\over N_\alpha !}\right]\simeq
-N\sum_\alpha\left[-1 +p(\alpha)\log
{p(\alpha)\over \delta}\right],
$$
where $p(\alpha) = N_\alpha/N, \delta = |\delta_\alpha| = 4^{-k}$, 
and we have used Stirling's
{}formula appropriate for $N_\alpha >> 1$.  Letting now $N\to \infty$
{}followed by $K\to \infty$ we obtain 
$$
N^{-1} S_B \to - \int^1_0\int^1_0 \bar f(x, y) \log \bar f (x, y) dxdy + 1
$$
where $\bar f(x, y)$ is the smoothed density, $p(\alpha) \sim \bar f
\delta$, which behaves according to the second law.  
In particular $p_t(\alpha)$ will approach the equilibrium state
corresponding to $p_{eq} (\alpha) = 4^{-k}$ while the empirical
density $f_t$ will
approach one in the unit square \cite{29}. N. B. If we had considered 
instead the Gibbs entropy $S_G $,\,  $ N^{-1} S_G =-\int_0^1\int^1_0 f_1 \log f_1
dx dy$, with $f_1(x, y$) the marginal, i.e. reduced, one particle distribution,
then this would not change with time. See section 7 and \cite{2}.

\section{Velocity Reversal}

The large number of atoms present in a macroscopic system plus the chaotic
nature of the dynamics ``of all realistic systems'' also explains why
it is so difficult, essentially impossible, for a clever
experimentalist to deliberately put such a system in a microstate which
will lead it to evolve in isolation, for any significant
time $\tau$, in a way contrary to the second law.\footnote{I
am not considering here entropy increase of the experimentalist and 
experimental apparatus directly associated with creating such a state.}
Such microstates certainly exist---just reverse all velocities Fig.\ 1b.  In fact, they are readily
created in the computer simulations with no round off errors, see Fig.\ 2 and 3.  To quote again from Thomson's article \cite{4}: ``If we allowed
this equalization to proceed for a certain time, and then reversed the
motions of all the molecules, we would observe a disequalization.  However,
if the number of molecules is very large, as it is in a gas, any slight
deviation from absolute precision in the reversal will greatly shorten the
time during which disequalization occurs.'' It is to be expected that
this time interval decreases with the increase of the chaoticity of
the dynamics.   In {\it addition}, if the
system is not perfectly isolated, as is always the case for real 
systems, the effect
of unavoidable small outside influences, which are unimportant for the
evolution of macrostates in which $|\Gamma_M|$ is increasing, will greatly
destabilize evolution in the opposite direction when the trajectory has to
{\it be aimed} at a very small region of the phase space.

The last statement is based on the very reasonable assumption that almost
any small outside perturbation of an ``untypical'' microstate 
$X\in \Gamma_{M(X)}$ will tend
to change it to a microstate $Y\in \Gamma_{M(X)}$ whose future time
evolution is typical of $\Gamma_{M(X)}$, i.e. $Y$ will likely be a
typical point in $\Gamma_{M(X)}$ so that typical behavior is not affected
by the perturbation \cite{14}.  If however we are, as in Figure 1, in a
micro-state $X_b$ at time $t_b$, where $X_b = T_\tau  X_a, \tau = t_b - t_a
> 0$, with $|\Gamma_{M_b}| >> |\Gamma_{M_a}|$, and we now reverse all
velocities, then $RX_b$ will be heading towards a smaller phase space 
volume during the
interval $(t_b, t_b+\tau)$ and this behavior is very untypical of
$\Gamma_{M_b}$. The velocity reversal therefore requires ``perfect aiming''
and will, as noted by Thomson \cite{4}, very likely be derailed by even small imperfections in the
reversal and/or tiny outside influences.  After a {\it very short} time
interval $\tau^\prime << \tau$, in which $S_B$ decreases, the imperfections
in the reversal and the ``outside'' perturbations, such as those coming from
a sun flare, a star quake in a distant galaxy (a long time ago) or from a
butterfly beating its wings \cite{14}, will make it increase again. This is
clearly illustrated in Fig.\ 3, which shows how a small perturbation
has no effect on the forward macro evolution, but completely destroys the time
reversed evolution.

The situation is somewhat analogous to those pinball
machine type puzzles where one is supposed to get a small metal ball into a
small hole.  You have to do things just right to get it in but almost
any vibration gets it out into larger regions.  For the macroscopic
systems we are considering, the disparity between relative sizes of the
comparable regions in the phase space is unimaginably larger$^e$ than
in the puzzle, as noted in
the example in Section 1.  In the absence of any ``grand conspiracy'', the
behavior of such systems can therefore be confidently predicted to be in
accordance with the second law (except possibly for very short time
intervals).  This is the reason why even in those special cases such as
spin-echo type experiments where the creation of an effective $RT_\tau
X$ is possible, the
``anti-second law'' trajectory lasts only for a short time \cite{30}.  In
addition the {\it total} entropy change in the whole process, including
that in the apparatus used to affect the spin reversal, is always positive in
accord with the second law.

\section{Cosmological Considerations}

Let us return now to the big question posed earlier: what is special about
$t_a$ in Fig.\ 1 compared to $t_b$ in a world with symmetric
microscopic laws?  Put differently,
where ultimately do initial conditions, such as those assumed at $t_a$, come
{}from?  In thinking about this we are led more or less inevitably to
introduce cosmological considerations by postulating an initial
``macrostate of the universe'' having a very small Boltzmann entropy, 
(see also 1$e$ and 1$f$).    To again quote Boltzmann \cite{31}: ``That in nature the transition
{}from a probable to an improbable state does not take place as often as the
converse, can be explained by assuming a very improbable [small $S_B$]
initial state of the entire universe surrounding us.  This is a reasonable
assumption to make, since it enables us to explain the facts of experience,
and one should not expect to be able to deduce it from anything more
{}fundamental''.  While this requires that the initial macrostate of the
universe, call it $M_0$, be very far from equilibrium with
$|\Gamma_{M_0}|<< |\Gamma_{M_{eq}}|$, it does not require that we
choose a special microstate in $\Gamma_{M_0}$.  As also noted by
Boltzmann elsewhere
``we do not have to assume a special type of initial condition in
order to give a mechanical proof of the second law, if we are willing to
accept a statistical viewpoint...if the initial state is chosen at
random...entropy is almost certain to increase.'' 
This is a very important aspect of Boltzmann's insight, it is
sufficient 
to assume that this microstate
is typical of an initial macrostate $M_0$ which is far from equilibrium.

This going back to the initial conditions, i.e. the existence of an
early state of the universe (presumably close to the big bang) with a
much lower value of $S_B$ than the present universe, as an
ingredient in the
explanation of the observed time asymmetric behavior, bothers some
physicists.  It really shouldn't since the initial state of the
universe plus the dynamics determines what is happening at present.
Conversely, we can deduce information about the initial state from what
we observe now.  As put by Feynman \cite{15},
 ``it is necessary to
add to the physical laws the hypothesis that in the past the universe was
more ordered, in the technical sense, [i.e. low $S_B$] than it is
today...to make an understanding of the irreversibility.''  A very
clear discussion of this is given by Roger Penrose in connection with
the ``big bang''
cosmology \cite{16}.  He takes for the initial macrostate of the universe
the smooth energy density state prevalent soon after the big bang: an
equilibrium state (at a very high temperature) {\bf except} for the
gravitational degrees of freedom which were totally out of
equilibrium, as evidenced by the fact that the matter-energy density was
spatially very uniform.  That such a uniform density corresponds to a
nonequilibrium state may seem at first surprising, but gravity, being purely attractive and long range, is
unlike any of the other fundamental forces.  When there is enough matter/energy
around, it completely overcomes the tendency towards uniformization
observed in ordinary objects at high energy densities or temperatures.
Hence, in a universe dominated, like ours, by gravity, a uniform density
corresponds to a state of very low entropy, or phase space volume, for a
given total energy, see Fig.\ 4.

\epsfysize=2in
\begin{figure}[t]
\centerline{\epsffile{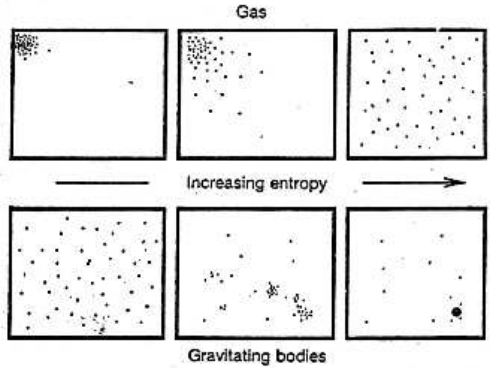}}
\caption{\it With a gas in a box, the maximum entropy state
(thermal equilibrium) has the gas distributed uniformly; however, with a
system of gravitating bodies, entropy can be increased from the uniform
state by gravitational clumping leading eventually to a black hole.  From
Ref.\ \cite{16}.} 
\end{figure}  

The local `order' or low entropy we see around us (and elsewhere)---from
complex molecules to trees to the brains of experimentalists preparing
macrostates---is perfectly consistent with (and possibly even a necessary
consequence
of, i.e.\ typical of) this initial macrostate of the universe.  The value of
$S_B$ at the present time, $t_p$, corresponding to 
$S_B (M_{t_p})$ of our current clumpy
macrostate describing a universe of planets, stars, galaxies, and
black holes, is much much larger than $S_B(M_0)$, the Boltzmann entropy of the
``initial state'', but still
quite far away from $S_B(M_{eq})$
its equilibrium value.  The `natural' or `equilibrium'
state of the universe, $M_{eq}$,
is, according to Penrose \cite{16}, one with all matter and
energy collapsed into one big black hole.  Penrose gives an estimate
$S_B(M_0) / S_B(M_{t_p}) / S_{eq}\sim 10^{88} / 10^{101} / 10^{123}$ in
natural (Planck) units, see Fig.\ 5.  (So we
may still have a long way to go.)

\epsfysize=2in
\begin{figure}[t]
\centerline{\epsffile{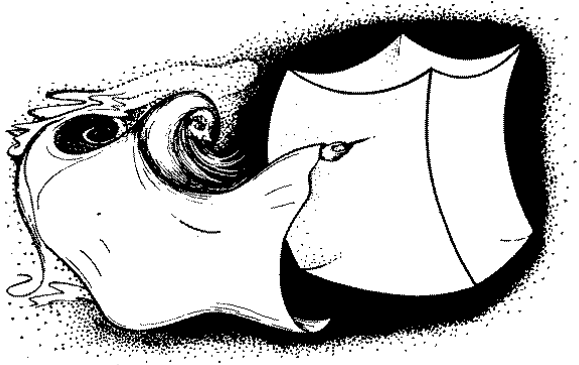}}
\caption{\it The creator locating the tiny region of
phase-space---one part in $10^{10^{123}}$---needed to produce a
$10^{80}$-baryon closed universe with a second law of thermodynamics in the
{}form we know it.  From Ref. \cite{16}. If the initial state was chosen
randomly it would, with overwhelming probability, have led to a universe in a
state with maximal entropy.  In such a universe there would be no
stars, planets, people or a second law.}
\end{figure}

I find Penrose's consideration about the very far from equilibrium
uniform density initial state
quite plausible, but it is obviously far from proven.   In any case it is, as
{}Feynman says, both
necessary and sufficient to assume a far from equilibrium initial
state of the universe, and this is in accord with all cosmological evidence.  
The question as to why the universe started 
out in such a very
unusual low entropy initial state worries R. Penrose quite a lot
(since it is not explained by any current theory) but such
a state is just accepted as a given by Boltzmann.   My own feelings are in
between.  It would certainly be nice to have a theory which would
explain the ``cosmological initial state'' but I am not holding my
breath.  Of course, if one believes in the ``anthopic principle'' in
which there are many universes and ours just happens to be right or we
would not be here then there is no problem -- but I don't find this
very convincing \cite{32}.

\section{Boltzmann vs. Gibbs Entropies}

The Boltzmannian approach, which focuses on the evolution of a single
macroscopic system, is conceptually different from what is commonly
referred to as the Gibbsian approach,
which focuses primarily on probability distributions or ensembles. This difference shows up strikingly
when we compare Boltzmann's entropy---defined in (1) for a microstate $X$
of a macroscopic system---with the more commonly used (and misused) entropy
$S_G$ of Gibbs, defined for an ensemble density $\rho(X)$ by
$$
S_G (\{\rho \}) = -k {\textstyle \int} \rho (X) [\log \rho(X)]dX.
\eqno (3)
$$

\noindent Here $\rho(X)dX$ is the probability (obtained some way or other)
{}for the microscopic state of the system to be found in the phase space
volume element $dX$ and the integral is over the phase space $\Gamma$.
Of course if we take $\rho(X)$ to be the generalized microcanonical ensemble
associated with a macrostate $M$,
$$\rho_M(X) \equiv \left \{
\matrix {
| \Gamma_M |^{-1}, & {\rm if}\ X \in \Gamma_M \cr
0, \hfill & {\rm otherwise}\hfill \cr
} \right. ,\eqno (4)\, 
$$
then clearly,
$$
S_G(\{\rho_M \}) = k\log |\Gamma_M | = S_B(M). \eqno (5)
$$

The probability density $\rho_{M_{eq}}$ for a system in the equilibrium
macrostate $M_{eq}$ is, as
already noted, essentially the same as that for the microcanonical (and
equivalent also to the canonical or grandcanonical) 
ensemble when the system is of macroscopic size.  Generalized
microcanonical ensembles $\rho_M(X)$, or their canonical versions,
are also often used to describe systems in which the particle density, energy
density and momentum density vary slowly on a microscopic scale {\it and}
the system is, in each small macroscopic region, in equilibrium with the
prescribed local densities, i.e.\ in which we have local equilibrium \cite{18}.  In
such cases $S_G(\{\rho_M\})$ and $S_B(M)$ agree with each other, and with
the macroscopic hydrodynamic entropy, to leading order in system
size$^d$. (The $\rho_M$ do not however describe the fluxes in such
systems: the average of $J(X)$, the microscopic flux function, being
zero for $\rho_M$ \cite{18}.)

The time evolutions of $S_B$
and $S_G$ subsequent to some initial time when $\rho = \rho_M$ are
{\it very} different, unless $M= M_{eq}$ when there is no further
systematic change in $M$ or $\rho$.
As is well known, it follows from the
{}fact (Liouville's thoerem) that the volume of phase space regions remains unchanged under the
Hamiltonian time evolution (even though their shape changes greatly) that
$S_G(\{ \rho \})$ never changes in time as long as $X$ evolves according to
the Hamiltonian evolution, i.e.\ $\rho$ evolves according to the Liouville
equation.  $S_B(M)$, on the other hand, certainly does change. Thus, if we
consider the evolution of the microcanonical ensemble corresponding to the
macrostate $M_a$ in Fig.\ 1a after removal of the constraint, $S_G$ would
equal $S_B$ initially but subsequently $S_B$ would increase while $S_G$
would remain constant. $S_G$ therefore does not give any indication that
the system is evolving towards equilibrium.

This reflects the fact, discussed earlier, that the probabililty
density $\rho_t(X)$ does not remain uniform over the domain
corresponding to the macrostate $M_t= M(X_t)$ for $t > 0$.  I am
thinking here of the case in which $M_t$ evolves deterministicaly 
so that almost all $X$ initially in
$\Gamma_{M_0}$ will be in $\Gamma_{M_t}$ at time $t$, c.f. \cite{2}.  As
long as the system remains truly isolated $\rho_t(X)$ will contain
memories of the initial $\rho_0$ in the higher order correlations, 
which are reflected in the complicated shape which an
initial region $\Gamma_{M_0}$ takes on in time but which do not affect the
{}future time evolution of $M$ (see the discussion at end of section 4).  
{\it Thus the relevant entropy for understanding the
time evolution of macroscopic systems is $S_B$ and not $S_G$}.  

Of course, if we do, at each time $t$,  a ``coarse graining'' of $\rho$ over the cell $\Gamma_{M_t}$
then we are essentially back to dealing with $\rho_{M_t}$, and we are just defining $S_B$ in a backhanded way.  This
is one of the standard ways, used in many textbooks, of reconciling the
constancy of $S_G$ with the behavior of the entropy in
real systems.  I fail to see what
is gained by this except to obscure the fact that the microstate of
a system is specified at any instant of time by a single phase
point $X_t$ and that its evolution in time is totally independent of how
well we know the actual value of $X_t$.  Why not use $S_B$ from the beginning?  We can of
course still use ensembles for computations.  They will yield correct
results whenever the quantities measured, which may involve averages
over some time interval $\tau$, have small dispersion in 
the ensemble considered.

\section{Quantitative Macroscopic Evolution}

Let me now describe briefly the very important but very daunting task
of actually rigorously deriving time asymmetric hydrodynamic equations from
reversible microscopic laws \cite{18}, \cite{19}.  While many qualitative features of
irreversible macroscopic behavior depend very little on the positivity of
Lyapunov exponents, ergodicity, or mixing properties of the microscopic
dynamics, such properties are important for the existence of a
quantitative description
of the macroscopic evolution via time-asymmetric
{\it autonomous} equations of hydrodynamic type.  The existence and form of
such equations depend on the rapid decay of correlations in space and
time, which requires chaotic dynamics. When the chaoticity can be
proven to be strong
enough (and of the right form) such equations can be derived rigorously
{}from the reversible microscopic dynamics by taking limits in which the
ratio of macroscopic to microscopic scales goes to infinity.  Using the law
of large numbers one shows that these equations describe the behavior of
almost all individual systems in the ensemble, not just that of ensemble
averages, i.e.\ that the dispersion goes to zero in the scaling limit.  The
equations also hold, to a high accuracy, when the macro/micro ratio is
{}finite but very large \cite{18}.

As already mentioned, an example in which this can be worked out in detail is the periodic
Lorentz gas.  This consists of a {\it macroscopic
number of non-interacting particles} moving among a periodic array of fixed
convex scatterers, arranged in the plane in such a way that there is a
maximum distance a particle can travel between collisions (finite
horizon Sinai billiard). The chaotic
nature of the microscopic dynamics, which leads on microscopic time
scales to an approximately
isotropic local distribution of velocities, is directly responsible for the
existence of a simple autonomous deterministic description, via a diffusion
equation, for the macroscopic particle density profile of this system 
\cite{18}.  A
second example is a system of hard spheres at very low densities for which
the Boltzmann equation has been shown to describe (at least for short
times) \cite{17} the evolution of 
$f_t ({\bf r}, {\bf v})$, the empirical density in the six dimensional position and velocity space.  I use these examples, despite their highly idealized
nature, because here (and unfortunately only here) all the mathematical i's have been dotted.  They thus
show {\it ipso facto}, in a way that should convince even (as Mark Kac put
it) an ``unreasonable" person, not only that there is no conflict between
reversible microscopic and irreversible macroscopic behavior but also
that, in these cases at least,
{\it for almost all initial microscopic states consistent with a given
nonequilibrium macroscopic state}, the latter follows from the former---in
complete accord with Boltzmann's ideas.

\section{Quantum Mechanics}

While the above analysis was done, following Maxwell, Thomson and
Boltzmann, in terms of classical mechanics, the
situation is in many ways similar in quantum mechanics.  Formally the
reversible incompressible flow in phase space is replaced by the unitary
evolution of wave functions in Hilbert space and velocity reversal of $X$
by complex conjugation of the wavefunction $\Psi$.  In particular, 
I do not believe that
quantum {\it measurement} is a {\it new} source of irreversibility.
Rather, real measurements on quantum systems are time-asymmetric because
they involve, of necessity, systems with a very large number of degrees of
{}freedom whose irreversibility can be understood using natural extensions of
classical ideas \cite{33}.  There are however also some genuinely new features in
quantum mechanics relevant to our problem; for a more complete
discussion see \cite{34}.

\medskip
\noindent{\bf Similarities}
\medskip

Let me begin with the similarities.  The analogue of the Gibbs entropy
of an ensemble, Eq (3), is the well known von Neumann entropy of a
density matrix $\hat \mu$,
$$
\hat S_{vN}(\hat\mu) = - k T r 
 \hat \mu \log \hat \mu.
\eqno (6)
$$ 
This entropy, like the classical $S_G(\rho)$, does not change in time
{}for an isolated system evolving under the Schr\"odinger time evolution
\cite{13}, \cite{35}.  Furthermore it has the value zero whenever 
$\hat \mu$ represents a pure state.  It is thus, like $S_G (\rho)$, not appropriate for describing the time
asymmetric behavior of an isolated macroscopic system.  We therefore naturally
look for the analog of the Boltzmann entropy given by Eq (1).
We shall see that  while the quantum version of (1b) is straight forward there
is no strict analog of (1a) with $X$ replaced by $\Psi$ which holds for all $\Psi$.

In a surprisingly little quoted part
of his famous book on quantum mechanics \cite{35}, von Neumann discusses 
what he calls the macroscopic entropy of a system.
To begin with, von
Neumann describes a macrostate $M$ of a macroscopic
quantum system\footnote{von Neumann unfortunately does
not always make a clear distinction between systems of macroscopic
size and those consisting of only a few particles and this leads I
believe to much confusion, c.f. article by Kemble \cite{36}.  See also articles
by Bocchieri and Loinger \cite{37}, who say it correctly. } by specifying the values of a set of
``rounded off'' commuting macroscopic observables, i.e. operators $\hat M$,
representing particle number, energy, etc.,\ in each of the cells into which
the box containing the system is divided (within some tolerence).  Labeling the
set of eigenvalues of the $\hat M$ by $M_\alpha$, $\alpha =
1,...$, one then has, corresponding to the collection $\{M_\alpha\}$,
an orthogonal decomposition of the system's Hilbert space $\cal H$ into linear
subspaces $\hat\Gamma_\alpha$ in which the observables $\hat M$ take the
values of $M_\alpha$.  (We use here the subscript $\alpha$ to avoid
confusion with the operators $\hat M$.)

Calling $E_\alpha$ the projection
into $\hat\Gamma_\alpha$, von Neumann then defines the {\it
macroscopic entropy} of a system with a density matrix; $\hat\mu$ as,
$$
\hat S_{mac}(\hat \mu) =k \sum^L_{\alpha=1} p_\alpha(\hat\mu) \log |\hat \Gamma_\alpha|
- k \sum^L_{\alpha=1} p_\alpha(\hat\mu) \log p_\alpha(\hat\mu) \eqno(7)
$$
where $p_\alpha(\hat \mu)$ is the probability of finding the system with
density matrix $\hat \mu$ in the macrostate $M_\alpha$,
$$
p_\alpha (\hat\mu) = Tr (E_\alpha \hat\mu),     \eqno(8)
$$
and $|\hat \Gamma_\alpha|$ is the dimension of $\hat\Gamma_\alpha$ (Eq
(6) is at the bottom of p 411 in \cite{35}; see also Eq (4) in \cite{36}).
An entirely analogous definition is made for a system represented by a
wavefunction $\Psi$: we simply replace $p_\alpha(\hat\mu)$ in (7) and
(8) by $p_\alpha(\Psi) = (\Psi, E_\alpha \Psi)$.  In fact $| \Psi > <
\Psi|$ just corresponds, as is well known, to a particular ({\it pure}) density matrix
$\hat \mu$.

Von Neumann justifies (7) by noting that 
$$
\hat S_{\textstyle mac }(\hat \mu) =-k  Tr[\tilde \mu  \log\tilde \mu]=S_{vN}(\tilde\mu)  \eqno(9)
$$
{}for  
$$
\tilde\mu = \sum(p_\alpha / |\hat \Gamma_\alpha|)E_\alpha \eqno(10)
$$
and that $\tilde \mu$ is macroscopically indistinguishable from $\hat \mu$.
This is analogous to the classical ``coarse graining'' discussed at the end
of section 7, with $\tilde\mu_\alpha = E_\alpha /|\hat \Gamma_\alpha| \leftrightarrow \rho_{M_\alpha}$
there.

It seems natural to make the correspondence between the
partitioning of classical phase space $\Gamma$ and the decomposition of the
Hilbert space $\cal H$ and to define the natural quantum analogue to Boltzmann's definition of $S_B(M)$ in
(1), as
$$
\hat S_B(M_\alpha) = k \log |\hat \Gamma_{M_{\alpha}}|\eqno(11)
$$
where $|\hat \Gamma_M|$ is the dimension of $\hat \Gamma_M$.  This is
in fact done more or less explicitely in \cite{35}, \cite{37},
\cite{13}, \cite{38} and is clearly consistent with the standard 
prescription for computing the von Neumann quantum
entropy of an equilibrium systems, with 
$\hat \mu = \hat \mu_{eq}$, where $\hat\mu_{eq}$ is the microcanonical
density matrix; $\hat \mu_{eq}\sim \tilde\mu_\alpha$, corresponding to
$M_\alpha = M_{eq}$.  
This was in fact probably standard at one time but since
{}forgotten.  In this case the right side of (11) is, to
leading order in the size of the system, equal to the von Neumann entropy
computed from the microcanonical density matrix ({\it as it is for
classical systems}) \cite{13}.

With this definition of $\hat S_B(M)$, the first term on the
right side of equation (7) is just what we would intuitively write down for
the expected value of the entropy of a classical system of whose macrostate
we were unsure, e.g.\ if we saw a pot of water on the burner and made some
guess, described by the probability distribution $p_\alpha$, about its
temperature or energy.  The second term in (7) will
be negligible compared to the first term for a macroscopic system, classical
or quantum, going to zero when divided by the number of particles in the
system.

One can give arguments for expecting $\hat S_B(M_t)$ to increase (or
stay constant) with $t$ after a constraint is lifted in a macroscopic
system until the system reaches the macrostate $M_{eq}$ \cite{38}. \cite{37},
\cite{34}.  
These arguments
are on the heuristic conceptual level analogous to those given above for
classical systems, although there are at present no worked out
examples analogous to those described in the last section.  This will
hopefully be remedied in the near future.  

\medskip
\noindent{\bf Differences}
\medskip

We come now to the differences between the classical and quantum pictures.
While in the classical case the actual state of the system is
described by $X\in \Gamma_\alpha$, for some
$\alpha$, so that the system is always definitely in one of the
macrostates $M_\alpha$, this is not so for a quantum system  specified
by $\hat\mu$ or $\Psi$.  We thus do not have the analog of (1a) for
general $\hat\mu$ or $\Psi$.    In fact, even when the system is in a
microstate $\hat\mu$ or $\Psi$ corresponding to a definite macrostate at
time $t_0$, only a classical system will always be in a unique
macrostate for all times $t$.  The quantum system may evolve to a
superposition of different macrostates, as happens in the well known
Schr{\"o}dinger cat paradox:  a wave function $\Psi$
corresponding to a particular macrostate evolves into a linear combination of
wavefunctions associated with very different macrostates, one corresponding
to a live and one to a dead cat (see references \cite{38} - \cite{41}).  

The possibility of superposition of wavefunctions is of course a general,
one might say the central, feature of quantum mechanics.  It is reflected
here by the fact that whereas the relevant classical phase space can be
partitioned into cells $\Gamma_M$ such that every $X$ belongs to exactly
one cell, i.e. every microstate corresponds to a unique macrostate, this is
not so in quantum mechanics.  The superposition principle rules out any
such meaningful partition of the Hilbert space: all we have is an
orthogonal decomposition.  Thus one cannot associate a definite macroscopic
state to an arbitrary wave function of the system.  This in turn raises
questions about the connection between the quantum formalism and our
picture of reality, questions which are very much part of the fundamental
issues concerning the interpretation of quantum mechanics as a theory of
events in the real world; see \cite{16}, and \cite{38}--\cite{44}
and references there for a discussion of these problems.

Another related difference between classical and quantum mechanics is that
quantum correlations between separated systems arising from wave function
entanglement render very problematic, in general, our assigning a wave 
{}function to a
subsystem ${\cal S}_1$ of a system ${\cal S}$ consisting of parts 
${\cal S}_1$, and ${\cal S}_2$ 
even when there is no direct interaction between ${\cal S}_1$ and
${\cal S}_2$.  This makes the standard idealization of physics---
an isolated system---much more problematical in quantum mechanics than in
classical theory.  In fact any system, considered
as a subsystem of the universe described by some wavefunction $\Psi$, will
in general not be described by a wavefunction but by a density matrix,
$\mu^\Psi = Tr|\Psi > < \Psi |$ where the trace is over ${\cal S}_2$.  

It turns out that for a small system coupled to a large system, which
may be considered as a heat bath, the density matrix of the small
system will be the canonical one, $\hat \mu_s = \hat \mu_\beta \sim
\exp [-\beta\hat H_s]$ \cite{45}, \cite{46}.  To be more precise,
assume that the (small) system plus heat bath $(s + B)$ are described
by a microcanonical ensemble, specified by giving a uniform
distribution over all normalized wave functions $\Psi$ of $(s + B)$ in
an energy shell $(E, E + \delta E)$.   Then the reduced density matrix of
the system $\hat\mu_s = Tr_B |\Psi ><\Psi |$ obtained from any typical 
$\Psi$ will be close
to $\hat\mu_\beta$, i.e. the difference between them will go to zero (in
the trace norm) as the number of degrees of freedom in the bath goes
to infinity.  This is a remarkable property of quantum systems which 
has no classical
analogue.  All one can say classically is that if one averges over the
microstates of the bath one gets the canonical Gibbs distribution for
the system.  This is of course also true and well known for quantum
systems, but what is new is that this is
actually true for almost all pure states $\Psi$, \cite{45}, \cite{46} see also references there to
earlier work in that direction, including \cite{37}.

One can even go further and find the dsitribution of the ``wave
{}function'' $\varphi$ of the small system described by 
$\hat\mu_\beta$ \cite{47}.
{}For ways of giving meaning to the wavefunction of a subsystem, see \cite{41} -
\cite{42} and \cite{48}.

\section{Final Remarks}

As I stated in the beginning, I have here completely ignored relativity,
special or general.  The phenomenon we wish to explain, namely the
time-asymmetric behavior of macroscopic objects, has
certainly many aspects which are the same in the relativistic (real)
universe as in a (model) non-relativistic one.  The situation is of
course very different when we consider the entropy of black holes and 
the nature of the appropriate
initial cosmological state where relativity is crucial.  Similarly the
questions about the nature of time mentioned in the beginning of this
article cannot be discussed meaningfully without relativity.  Such
considerations may yet lead to entirely different pictures of the
nature of reality and may
shed light on the
interpretation of quantum mechanics, discussed in the last section,
c.f. \cite{16}.  Still it is
my belief that one can and in fact one must, in order to make any scientific
progress, isolate segments of reality for separate analysis.  It is only
after the individual parts are understood, on their own terms, that one can
hope to synthesize a {\it complete picture}.

To conclude, I believe that the Maxwell-Thomson-Boltzmann resolution of the
problem of the origin of macroscopic irreversibility contains, in the
simplest idealized classical context, the essential ingredients for
understanding this phenomena in real systems.  Abandoning Boltzmann's
insights would, as Schr{\"o}dinger says\footnote{Schr{\"o}dinger 
writes \cite{26}, ``the spontaneous transition from order to
disorder is the quintessence of Boltzmann's theory \
\dots \ This theory really grants an understanding and does not \
\dots \ reason away the dissymmetry of things by means of an a priori
sense of direction of time variables\dots  \ \ No one who has once
understood Boltzmann's theory will ever again have recourse to
such expedients.  It would be a scientific regression beside which a
repudiation of Copernicus in favor of Ptolemy would seem trifling.''}
be a most serious scientific
regression.  I have yet to see any good reason to doubt Schr{\"o}dinger's
assessment.
\medskip

\noindent
{\bf Acknowledgments.}  I greatfully acknowledge very extensive and very
useful discussions with my colleagues Sheldon Goldstein and Eugene
Speer.   I also thank Giovanni Gallavotti, Oliver Penrose and David
Ruelle for many enlightening discussions and arguments.  I am also
much indebted to S. Goldstein, O. Penrose and E. Speer for careful
reading of the manuscript.
Research supported in part by NSF Grant DMR 01279-26 and AFOSR Grant AF-FA9550-04.

\end{document}